\documentclass{article}
 \usepackage{graphicx}
\title{Interacting Many-Investor Models, Opinion Formation and Price Formation with Non-extensive Statistics.}
\author{Fredrick Michael* \\ \it{written 01/2001, revised 11/2008}} 
\begin{document}
\maketitle
\abstract{We seek to utilize the nonextensive statistics to the microscopic modeling of 
the interacting many-investor dynamics that drive the price changes in a market. The statistics of price changes are known to be fit well by the Students-T and power-law distributions of the nonextensive statistics. We therefore derive models of interacting investors that are based on the nonextensive statistics and which describe the excess demand and formation of price.}

\section{Introduction}
As  is known from  economics, the price  of a security  (for example) can  be
related to the law of supply and demand. That is to say, the excess demand
is  proportional to the price such that we can write approximately
\begin{equation}
\includegraphics[width=40mm]{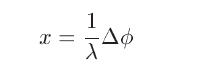} \label{eqn1}
\end{equation}

and here $\lambda$ is the  market depth \cite{4,5}.
In the past decade or so, there have been  many models  \cite{4,5} proposed that
attempt to  capture the dynamics and  statistics of market participants.  These
range from minority game models \cite{11}, multi-agent models, and lattice super-spin models that encode the many degrees of freedom available to an interacting investor as the degrees of freedom of the variables and spins of the models. These models attempt to quantify the
excess demand brought about by the mismatch of supply and demand between
interacting investors in a  market. The hallmark for the success of a model
has been the ability of the model in reproducing the stylized facts of real
markets. These are the heavy tails  (power-law) of the distributions, anomalous
(super) diffusion, and therefore statistical dependence (long-range correlations)
of subsequent price changes.

Recently we reported on an application of the C.Tsallis nonextensive statistics to the S$\&$P500 stock index \cite{1,3}. There we argued that the statistics are applicable
to a broad range of  markets and exchanges where anamolous (super) diffusion
and 'heavy' tails of the distribution are present, as  they are in the S$\&$P500 \cite{3}.
In effect we have characterized the statistics of the price changes (the left hand
side  of Eq.(1) ) as being well-modeled  by the non-extensive statistics. We now
seek to examine  the demand-side of the equation in light of our recent findings
that the non-extensive statistics models well the statistics of the price changes in
real markets.  As such, we will seek to outline a method by which one can obtain
many-investor models within the context of the Tsallis nonextensive statistics.
We will  derive our  specific models  utilizing the well-known  techniques of  the
maximum entropy approach \cite{6,8}.
Let us briefly review the maximum entropy approach to be utilized here. The
nonextensive, least-biased probability distribution function (PDF) $P(z,t)$ of an
underlying observable $z(t)$ is obtained by maximizing an incomplete information
theoretic measure equivalent to the Tsallis entropy $S_q$ \cite{1,  6,  8}
\begin{equation}
\includegraphics[width=80mm]{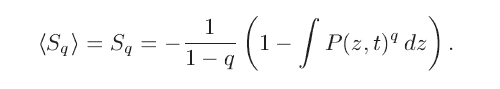} \label{eqn2}
\end{equation}
Here $P(z ,t)$  is the probability  distribution function  and will  be shown to
be  of a power-law form,  and is the degree of  non-extensivity or equivalently
the incompleteness  of  the information  measure.   The  inverse of the normalization is the partition function $Z(t)$,  and $\beta(t)$ is  a Lagrange  multiplier associated  with the
constraint(s).
In order  to build our  model(s), we must  specify the constraints  to be  utilized in our maximization procedure. These  constraints will be the  known (or
assumed) observables of interest, and  that are presumed to  capture the deterministic  behavior of our  many-investor system.  An  interesting model for  the
investors is  the model of  investor bias  and demand  developed  by Cont  and
Bouchaud and generalized to  super-spins by Chowdhury \cite{4,5,9}.  This model
assumes that the  generalized spins are representative of the magnitude and direction  of demand (the bias)  of the investors.  We will adopt this  model, with
suitable modifications, as a first approximation for characterizing the deterministic 
investor dynamics as observables from within the context of the nonextensive statistics.
Following the  work of Chowdhury et. al. let us define the  demand function
of a  system of interacting  investors as a classical  Hamiltonian-like function in
physics.   This Hamiltonian  will then  be  dependent on  the magnitude  of the
demand,  and the other degree  of freedom, the direction  of the demand  or the
bias.  The bias in  this model will  be taken in  this initial model  to be discrete
and will  be represented as a spin.  Also, for simplicity, let us initially assume
that the magnitude is  fixed.
The N-investor interaction potential in the Hamiltonian then  is taken to be
made of discrete terms with simple constant ferromagnetic coupling strengths
$(J_{ij} = J >0)$ and  an anti-ferromagnetic  coupling $ (L_{ij} =L >0)$ which we
treat in  the global mean field sense of the Bornholdt model \cite{9,10} such that
\begin{equation}
\includegraphics[width=100mm]{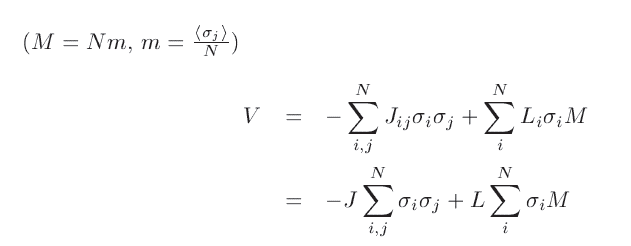} \label{eqn3}
\end{equation}

The  first  potential  term   can  be  seen   to  be   the  usual  form  of   a  ferromagnetic
spin-spin interaction with  a coupling  strength  and models the  herd-like, or
collective  opinion formation  of  investors.  The  spins  represent discrete  bias,
though the generalization to a continuous degree of bias is straightforward, and
can take on the values
\begin{equation}
\includegraphics[width=50mm]{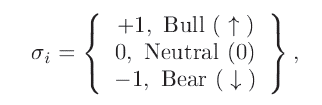} \label{eqn4}
\end{equation}
and  therefore when two  investors minimize their risk  and both  agree in their
bias $\sigma_i \sigma_j= 1$ the overall energy is  minimized. Note
that in this simple model if one of two investors is neutral, the interaction term is zero.
The second potential term is anti-ferromagnetic $(L>0)$  and models contrarian
investor behavior as in the Bornholdt model. We can simplify  the interactions
by assuming a mean field approximation.
Now let us assume that the bias 
fluctuates.  This is a reasonable assumption,
and  leads us to  the necessity of  describing the averages  of the 
fluctuations  in
terms of the mean and the variance.  These moments can be written for the $i$th
investor  as $q$-parametrized averages \cite{6,8}

\begin{equation}
\includegraphics[width=80mm]{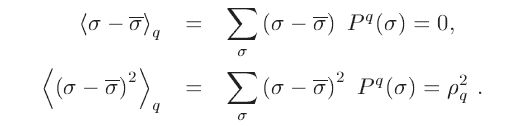} \label{eqn5}
\end{equation}

In  order  to  obtain a  tractable  solution  that  retains  the  essential  interaction
dynamics to lowest order, let us make a mean field approximation to our Hamiltonian.  This will allow us to linearize the spin-spin  interaction and will allow us to derive a 
closed form solution which, it is supposed, will allow us to qualify and approximately quantify the inherent
behavior of the system.  The full solution  to this model would perforce  involve
the complicated spin-spin terms and will result in corrections to our solved form
mean field solution.  As stated then, the mean field solution will give us approximately the first order response of  the system, given the interactions.  We then have ($H_{mf}$ is the mean field Hamiltonian)
\begin{equation}
\includegraphics[width=80mm]{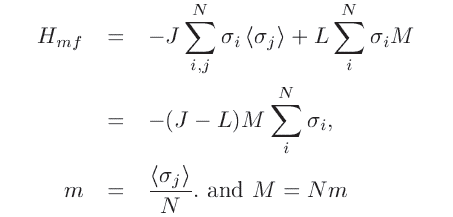} \label{eqn6}
\end{equation}
Here m is the  average bias per  investor and can  be seen  to be the  analog to
the  average magnetization per particle.  Next  let us build  into this model  the
open nature  of a market.   That is, the  number of  investors is not  a constant
in a market,  and we will  account for this fact  by the inclusion  of the number
of  investors as  a further  constraint (observable)  in our  model.   As such  the
observable (per investor) to be included in our  maximization procedure will be
\begin{equation}
\includegraphics[width=60mm]{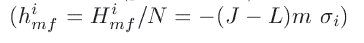} \label{eqn7}
\end{equation}
\begin{equation}
\includegraphics[width=30mm]{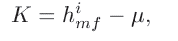} \label{eqn8}
\end{equation}

and $\mu$ is a total investor number $N$ constraint multiplier  which from the usual
thermodynamic  analogies goes  as the `chemical  potential'.  We  therefore can
write for the $i$th investor  the following  observables to  be included  in  our entropy
maximization (and dropping the $i$ sub- and superscripts)
\begin{equation}
\includegraphics[width=70mm]{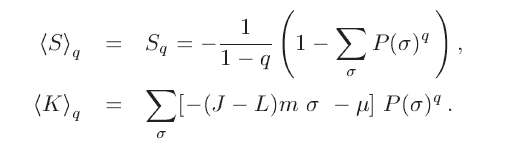} \label{eqn9}
\end{equation}

The  maximum  entropy  approach  then  allows  us   to  vary  the  entropy  given
the constraints such that
\begin{equation}
\includegraphics[width=40mm]{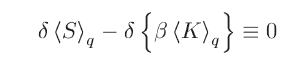} \label{eqn10}
\end{equation}
and  we  obtain   our  least  biased   probability  density  function   as  a  Tsallis   non-
extensive statistics power-law form
\begin{equation}
\includegraphics[width=70mm]{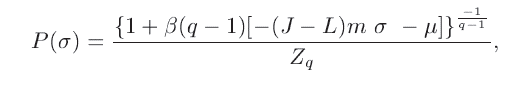} \label{eqn11}
\end{equation}
and  here  the  partition  function  is  related  to  the  normalization  and  is  given  by  $Z_q = \sum\limits{ \sigma}{}P(\sigma)$

We now wish  to examine  the  average bias in  this  model.  Following the  usual
magnetic systems argument, the average bias can  be written as (recall $\uparrow= +1, \downarrow =-1$)
\begin{equation}
\includegraphics[width=50mm]{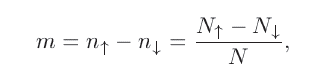} \label{eqn12}
\end{equation} 

The question  then is  how  to obtain $( N\uparrow, N\downarrow ) $ given that $N=N_{+}+ N_{0}+N_{-}$.
We can  write as before
\begin{equation}
\includegraphics[width=50mm]{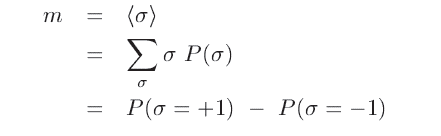} \label{eqn13}
\end{equation}
This expression  can then be related to the  average price  change by the market
depth and we obtain our desired result. That is, the Tsallis power-law statistical
distribution for  the price and price changes, as reported elsewhere \cite{1} for stock
market indices such as the S$\&$P500 high frequency price data, is obtained  from
the individual investor bias distributions with a proportionality factor of market
depth converting the  excess demand to price in  currency. We make use of  the
market depth $\lambda$ and write ($x$ is the price)
\begin{equation}
\includegraphics[width=30mm]{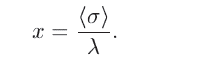} \label{eqn14}
\end{equation}
\section{continuous  spin  model}
We now wish to generalize beyond the limitations of the approximations we have
built into the pure spin model.  To do this, let us assume  that the state vector
for the system is  of two dimensions, the magnitude and the bias.  We will then
work with continuum spins  as in the Kosterlitz-Thouless  Hamiltonian.  Let us
write down the observables for  the interacting investors in the two dimensions
(y= magnitude,$\theta $=bias `angle')  of magnitude and bias as $V$. We then have

\begin{equation}
\includegraphics[width=80mm]{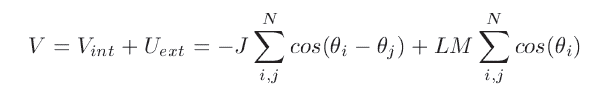} \label{eqn15gtha}
\end{equation}  

and  here $V_{int}$ is  the  Kosterlitz-Thouless interaction,  relegated  to  the role  of  the
potential. This total Hamiltonian with the inclusion of the previously discussed
total number $N$ of investors as a further observable will comprise the constraints
in the maximization of the entropy that we will perform  next. But first, let
us simplify the interaction term in the potential again by averaging over the $j$th
spins such  that  ($ \pi\le\theta\le 0$ here)
\begin{equation}
\includegraphics[width=90mm]{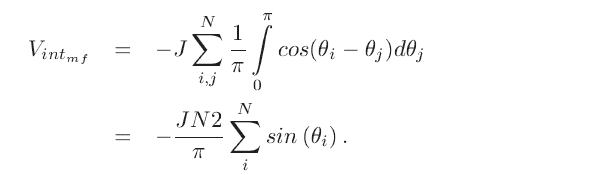} \label{eqn15gthb}
\end{equation}
again we can maximize the non-extensive entropy 
(per-investor) given the constraints of the  potential and  the first  two central moments  of the  
fluctuating variables of individual magnitude and  bias such that for the $i$th investor
we have
\begin{equation}
\includegraphics[width=90mm]{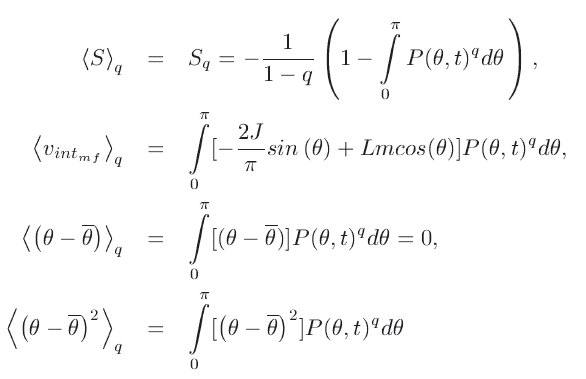} \label{eqn16}
\end{equation}

The  maximum  entropy method  then  states  that  we  must  maximize  the  entropy given the observables as  constraints.  This yields the following  variation
of the $q$-averaged observables

\begin{equation}
\includegraphics[width=100mm]{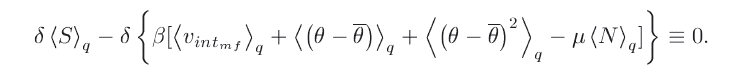} \label{eqn17}
\end{equation}

The least  biased PDF will  again be  of the  non-extensive form and  can be  written    
as
\begin{equation}
\includegraphics[width=110mm]{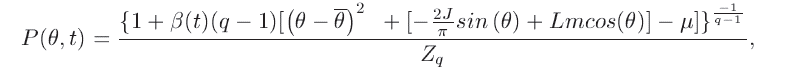} \label{eqn18}
\end{equation}
where the partition  function is again  related to the  normalization and is  now $Z_q =\int\limits_{0}{\pi}d\theta$

We   wish  to  obtain  the   average bias   and  relate  it  to   the  price  change.   We
then write the regular statistical average as $M$ and utilize the market depth  to
obtain the correspondence
\begin{equation}
\includegraphics[width=80mm]{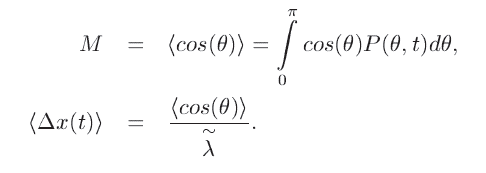} \label{eqn19}
\end{equation}

\section{continuous  model}
We now wish to relax all of the approximations and the simplifications we built into the previous spin  based models. To do this let us assume that the state vector for the system is comprised of two continuous degrees of freedom, the magnitude and the bias. Without restating the  problem, let  us write down  the
Hamiltonian for the investors in  the two dimensions (y=magnitude,$\theta$=bias
`angle')  of  magnitude and  bias  as $H_o$ (denotes   the   non-interacting   investor
Hamiltonian).  Also, let  us propose  some general  interaction potential $V$ the
form of which will be examined subsequently .  The important point here is that we are
seeking to cast the problem  of building a model for many  interacting-investors
into the powerful language of  the many-particle physics as  we feel this  allows
us to map some questions of  modeling financial markets and investor behavior
directly to well known physics-based paradigms. We have already touched upon
this in our specific models discussed above, and now we wish  to generalized the
application of the technique.

The model will consist of  the free and interacting parts of the Hamiltonian.  That is, the $i$th
investor will have $i(y(t),\sigma(t)) $ magnitude of demand  , with a bias of buy, sell or
hold, at time t.  The magnitude of demand can be assumed to be the demand for
number of  shares of a  stock, and with the  total number being  considered 
fixed as a  long-term constant $n=\int_{0}^{max}y(t)dy$.  The direction  of demand,  is

discrete $(-1,0,+1)$ corresponding to buy, sell or hold.  The Hamiltonian is
\begin{equation}
\includegraphics[width=60mm]{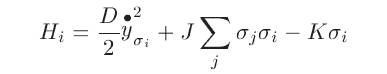} \label{eqn20}
\end{equation}

with the constraints of the moments
\begin{equation}
\includegraphics[width=80mm]{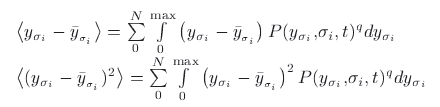} \label{eqn21b}
\end{equation}
the maximization of the entropy then obtains the least biased distribution
and with the normalization, $\sum \int P(\sigma_i, t)dy_{\sigma_i}=  1$ to  unity. 
The   extremization   yields   the   least   biased   probability   distribution   function, $P(\sigma_i, t)$
which upon taking the mean field approximation $\sum< \sigma_j>=M$ gives
\begin{equation}
\includegraphics[width=110mm]{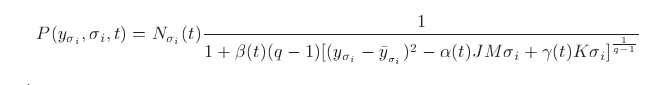} \label{eqn223b}
\end{equation}

The expectation value for the demand is the expression $< \sigma >=\sum_{\sigma,i}^{}\int { \sigma_i} P(y_{\sigma_i},\sigma_i , t) {dy_{\sigma_i}}$,
which   explicitly   is
\begin{equation}
\includegraphics[width=110mm]{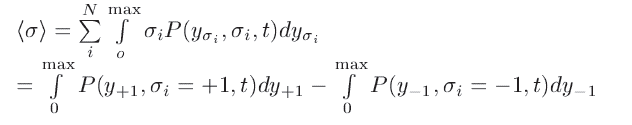} \label{eqn22}
\end{equation}

in  terms of  the  number  of shares  demanded  to  be  bought or  sold  or  held for
the moment at time t, the  expectation value of excess demand is

\begin{equation}
\includegraphics[width=60mm]{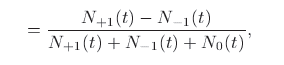} \label{eqn24c}
\end{equation}
This excess demand is proportional to the price and with the market depth $\lambda$ as the proportionality factor we obtain the price and change in price if the definition of the variables of excess demand is relative $y=x-x'$ 
\begin{equation}
\includegraphics[width=40mm]{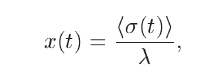} \label{eqn24}
\end{equation}
This model is perhaps the most detailed of the three discussed . It is also the most robust and points the way to the inclusion of interaction terms that describe interactions observed factors in real world markets. These could include the independent investor, institutional investor clustering, floor trader and outside trader time lags, non-constant market depths, more complex herding behavior etc. These can all be included as observables multiplied by Lagrange multipliers as constraints in the maximization, and the derived least biased distributions though perhaps complicated can be solved numerically.   
\section{conclusion}
The models presented here are of increasing complexity. However the three
models, discrete and continuous can be generalized further and can easily be applied. The numbers of shares  being bought, sold or held
at  any moment determine the instantaneous  excess demand.  This in turn is
related  to the instantaneous price or price change if relative variables are used, by the market  depth proportionality factor,  here assumed constant.  The
statistical distributions of price changes and price have been shown to be
well fitted  by the students-T distribution and more recently the  Tsallis nonex-
tensive statistics distribution . The information theoretic approaches taken here
assume the nonextensive entropy as a starting point, and  obtains not
surprisingly a power-law distribution  for the numbers of shares to bought, sold
or held.  These are  then summed and  the time-dependent price  change distributions obtained.   A question to  be answered  in subsequent work  is , what form of nonlinear interaction causes to arise a 
power-law distribution of price  changes  from a Gibbs-Boltzmann  form of extensive
entropy or information measure. How do nonlinear interactions modify the Gaussian distributions obtained from the extensive entropy.  Also, how does the numerical simulation  of the
model, by Monte Carlo or stochastic trajectory methods,  compare to actual market data price changes. Previous numerical simulations of power-law distributions have been shown by us to fit very well the high frequency stochastic time series data of price changes of the S$\&$P500 and we have also applied the power-law statistics to generalized Black-Scholes equations of prices for options and derivatives secondary markets for the underlying stocks and stocks indices etc.  We expect the present models especially the power-law distributed model to fit accurately the market price data , and given the parameters extracted from market data, the model can be a theory that begins to describe the statistical uncertainties, and on the average the microscopic interactions that occur between investors in a financial market and which then the theoretical model can subsequently be utilized for minimization of risk and the design of investment strategies in economics and finance.  
The  author  wishes to  state that  most  of this  manuscript was  written in
20  Nov 2001 with M.D.  Johnson and John  Evans at the University  of central
Florida Physics Department, Orlando, Florida. Recent research of the author's
has  refocused on this area of research, and the publication follows.
\pagebreak

\end{document}